\documentclass[conference]{IEEEtran}
\IEEEoverridecommandlockouts
\usepackage[utf8]{inputenc}
\usepackage{amsmath}
\usepackage{graphicx}
\usepackage{amsmath, amssymb}
\usepackage[printonlyused,nolist]{acronym}
\usepackage{xcolor}
\usepackage{algorithm}
\usepackage{algpseudocode}
\usepackage{etoolbox}
\usepackage{multirow}
\usepackage{tikz}
\usepackage{url}
\usepackage{graphicx}
\usepackage{tikz}
\usepackage{pgfplots}
\usepackage{optidef}
\usepackage{booktabs}
\usepackage{LMSnodeshapes}

\pgfplotsset{compat=1.15}

\usetikzlibrary{calc}

\DeclareMathOperator*{\w}{\boldsymbol{w}}
\DeclareMathOperator*{\X}{\boldsymbol{X}}
\DeclareMathOperator*{\bP}{\boldsymbol{P}}
\DeclareMathOperator*{\e}{\boldsymbol{e}}
\DeclareMathOperator*{\C}{\boldsymbol{C}}
\DeclareMathOperator*{\y}{\boldsymbol{y}}
\DeclareMathOperator*{\I}{\boldsymbol{I}}
\DeclareMathOperator*{\bs}{\boldsymbol{s}}
\DeclareMathOperator*{\herm}{\text{H}}

\begin{acronym}
	\acro{ENR}{echo-to-noise ratio}
	\acro{KF}{Kalman Filter}
	\acro{PBKF}{Partitioned-Block KF}
	\acro{PBAF}{Partitioned-Block Adaptive Filter}
	\acro{PBC}{Partitioned-Block Convolution}
	\acro{ML}{Maximum-Likelihood}
	\acro{EM}{Expectation-Maximization}
	\acro{AEC}{Acoustic echo cancellation}
	\acro{PF}{postfilter}
	\acro{GRU}{gated recurrent unit}
	\acro{DPF}{deep learning-based PF}
	\acro{MMSE}{Minimum Mean Square Error}
	\acro{STFT}{Short-Time Fourier Transform}
	\acro{PSD}{Power Spectral Density}
	\acro{ERLE}{Echo Return Loss Enhancement}
	\acro{RIR}{room impulse response}
	\acro{FIR}{Finite Impulse Response}
	\acro{PDF}{Probability Density Function}
	\acro{VSS}{Variable Step Size}
	\acro{DFT}{Discrete Fourier Transform}
	\acro{DNN}{Deep Neural Network}
	\acro{KL}{Kullback-Leibler}
	\acro{NER}{near-end-to-echo ratio}
	\acro{AF}{adaptive filter}
	\acro{EPC}{Echo Path Change}
	\acro{RTF}{Real-Time Factor}
\end{acronym}

\def\BibTeX{{\rm B\kern-.05em{\sc i\kern-.025em b}\kern-.08em
    T\kern-.1667em\lower.7ex\hbox{E}\kern-.125emX}}
\begin{document}

\title{A Synergistic Kalman- and Deep Postfiltering Approach to Acoustic Echo Cancellation \thanks{This work was supported by the German Research Foundation {- 282835863 -} within the Research Unit FOR2457 "Acoustic Sensor Networks.}}

\author{\IEEEauthorblockN{Thomas Haubner, Mhd. Modar Halimeh, Andreas Brendel, and Walter Kellermann}
	\IEEEauthorblockA{\textit{Multimedia Communications and Signal Processing,}
		\textit{Friedrich-Alexander-University Erlangen-N\"urnberg,}\\
		Cauerstr. 7, D-91058 Erlangen, Germany,
		e-mail: \texttt{thomas.haubner@fau.de}}}

%\title{Conference Paper Title*\\{\footnotesize \textsuperscript{*}Note: Sub-titles are not captured in Xplore andshould not be used}\thanks{Identify applicable funding agency here. If none, delete this.}}

%\name{Thomas Haubner, Mhd. Modar Halimeh, Andreas Brendel, and Walter Kellermann \thanks{This work was supported by the Deutsche Forschungsgemeinschaft (DFG, German Research Foundation) - 282835863 - within the Research Unit FOR2457 "Acoustic Sensor Networks.}} 
%\address{Multimedia Communications and Signal Processing, Friedrich-Alexander-University Erlangen-Nürnberg,\\	Cauerstr. 7, D-91058 Erlangen, Germany, Thomas.Haubner@FAU.de}
%\author{\IEEEauthorblockN{1\textsuperscript{st}Thomas Haubner}
%\and
%\IEEEauthorblockN{2\textsuperscript{nd} Mhd. Modar Halimeh}
%\and
%\IEEEauthorblockN{3\textsuperscript{rd} Andreas Brendel}
%\and
%\IEEEauthorblockN{3\textsuperscript{rd} Walter Kellermann}
%	\textit{name of organization (of Aff.)}\\
%	City, Country \\
%	email address or ORCID}
%}

\setlength{\textfloatsep}{0.35cm} % \setlength{\textfloatsep}{0.35cm}

\newcommand{\commentTHa}[1]{\textcolor{black}{#1}}
\newcommand{\commentTHb}[1]{\textcolor{black}{#1}}
\newcommand{\commentTHc}[1]{\textcolor{black}{#1}}
\newcommand{\commentTHd}[1]{\textcolor{black}{#1}}
\newcommand{\commentTHe}[1]{\textcolor{black}{#1}}		% you changed this last time
\newcommand{\commentTHf}[1]{\textcolor{black}{#1}}
\newcommand{\commentTHg}[1]{\textcolor{black}{#1}}
\newcommand{\commentTHj}[1]{\textcolor{black}{#1}}

\maketitle

\begin{abstract}
	% 140 words
	% We introduce a synergistic {adaptive Kalman filter} and neural network-based postfilter approach to double-talk robust acoustic echo cancellation in time-varying environments. 
	We introduce a synergistic approach to double-talk robust acoustic echo cancellation combining adaptive Kalman filtering with a deep neural network-based postfilter. The proposed algorithm overcomes the well-known limitations of Kalman filter-based adaptation control in scenarios characterized by abrupt echo path changes. As the key innovation, we suggest to exploit the different statistical properties of the interfering signal components for robustly estimating the adaptation step size. This is achieved by leveraging the postfilter near-end estimate and the estimation error of the Kalman filter. The \commentTHa{proposed scheme} allows for rapid reconvergence of the adaptive filter after abrupt echo path changes without compromising the steady-state performance achieved by state-of-the-art approaches in static scenarios.
	% The proposed synergistic scheme allows for rapid reconvergence of the adaptive filter after abrupt echo path changes without compromising the steady-state performance achieved by state-of-the-art approaches in static scenarios.
	%The proposed step size estimator leverages the postfilter near-end estimate and the adaptation error of the Kalman filter and allows for rapid reconvergence of the adaptive filter after abrupt echo path changes {without impairing the steady-state performance of }.
	%
	%{This is achieved by leveraging} the postfilter near-end estimate and the adaptation error of the Kalman filter. {The proposed algorithm efficiently uses the postfilter near-end estimate and the adaptation error }
	%
	% 
	%{without reducing the steady-state performance}
	% {By leveraging the postfilter near-end estimate and the adaptation error of the Kalman filter we exploit}
	%
	%The proposed algorithm represents a mutually supporting adaptive filter and post filter approach to acoustic echo cancellation in time-varying acoustic environments.
\end{abstract}

\section{Introduction}
\label{sec:intro} 
\ac{AEC} is an essential part of any full-duplex hands-free acoustic communication \commentTHd{application, e.g., teleconferencing or human-machine dialogue systems} \cite{enzner_acoustic_2014}. Research on \ac{AEC} algorithms has evolved from simple time-domain \commentTHb{\ac{AF}} approaches \cite{widrow_b_adaptive_1960} and computationally efficient frequency-domain implementations \cite{ferrara_fast_1980} to recent deep learning-based approaches \cite{zhang_deep_2019, westhausen2020acoustic}. 

In general, current \ac{AEC} algorithms can be classified into model-based \ac{AF}-\ac{PF} approaches and direct \ac{DPF} approaches. While model-based algorithms show unmatched generalization to unknown acoustic environments, they require sophisticated adaptation control mechanisms to cope with double-talk situations \cite{enzner_acoustic_2014}. In particular the probabilistically motivated inference of the \ac{AF} coefficients by a \ac{KF} 
%
% \cite{enzner_frequency-domain_2006, malik_online_2010, kuech_state-space_2014} 
\commentTHb{\cite{enzner_frequency-domain_2006, kuech_state-space_2014}} 
enabled the much sought-after continuous adaptation control without the need of a double-talk detector. However, despite the increased double-talk robustness, \ac{KF} approaches suffer from slow reconvergence after abrupt \acp{EPC} which are commonly encountered with portable devices \cite{yang_frequency-domain_2017, jiang_improved_2019}. This slow recovery is caused by overestimating the noise \ac{PSD} matrix and often remedied by auxiliary mechanisms like shadow filters \cite{yang_frequency-domain_2017} \commentTHb{or trained noise models \cite{kfNMF}} 
% \commentTHa{or trained noise models \cite{kfNMF}} 
which require additional computational power. 
% computational power and are notoriously difficult to parametrize
\commentTHa{Unlike \acp{AF}, \ac{DPF} algorithms for \ac{AEC} \cite{zhang_deep_2019, westhausen2020acoustic} do not require online adaptation when trained on adequate datasets. }
%\shorten{Thus, they do not need auxiliary mechanisms for handling interfering double-talk and \acp{EPC}.} 
\commentTHa{However, as communication devices are usually exposed to a large variety of acoustic environments, \commentTHb{\ac{DNN}} models with many parameters need to be trained which limits their applicability to devices with sufficient computational power.} These computational requirements can be mitigated by using smaller networks with less parameters when combining model-based and deep learning approaches \cite{Carbajal2018, pfeifenberger_nonlinear_2020, combAdFiltAndComValDPF}. However, most methods treat the \ac{AF} estimation independently from the \ac{DPF}. Recently, a \commentTHb{\ac{DNN}}-supported \ac{EM} optimization of a local Gaussian model has shown improved performance for joint reverberation, echo and noise reduction \cite{carbajal_joint_2020}. However, a narrowband assumption  is made and the filter estimates are assumed to be time-invariant which limits the performance for time-varying scenarios \cite{carbajal_joint_2020}. % \shorten{this motivation can be shortened}

%
%
%{In this paper, we introduce a computationally efficient and low-latency approach to \ac{AEC} by a synergistic combination of a broadband \ac{PBKF} and {a} {\ac{DPF}.}
%
In this paper, we introduce a synergistic approach which combines a broadband adaptive \ac{KF} and a \ac{DPF} and thereby successfully copes with time-varying acoustic environments. We show how the slow reconvergence of the \ac{KF} after abrupt \acp{EPC} can be remedied by exploiting the different signal statistics of the various interfering components. By efficiently fusing the \ac{DPF} near-end estimate and the \ac{KF} estimation error, a robust estimate for the \ac{KF} step size is obtained without any auxiliary mechanisms.
\begin{figure} [b] % TODO: maybe directly 
	\vspace*{-.17cm}
	\centering	
	\begin{tikzpicture}[node distance=1.5cm]
	\tikzset{loudspeaker style/.style={%
			draw,very thick,shape=loudspeaker,minimum size=5pt
	}}
	\tikzset{microphone style/.style={%
			draw,very thick,shape=microphone,minimum size=.5pt, inner sep=2.0pt
	}}
	
	\node (sigInX) at (0,0) {};
	\node [left of=sigInX, node distance=2cm, inner sep=.0] (sigInX1) {};
	\node [right of=sigInX, inner sep=0] (spltX) {};
	\node [below of=spltX, rectangle, draw, thick, node distance=1.0cm] (pbkf) {KF};
	\node [below of=pbkf, circle, draw, inner sep=2, thick, node distance=1.9cm] (subtraction) {};
	% \node [right of=subtraction] (sigInY) {};
	\node [rectangle, draw, thick] (pf)  at ($(subtraction)+(-4.0,.1)$) {PF};
	% \node [left of=subtraction, rectangle, draw, thick, node distance=3.25cm] (pf)  at ($(subtraction)+(-2,0)$) {PF};
	\node [left of=pf, thick] (epf) {};
	\node [left of=pbkf, rectangle, draw, node distance=2.2cm] (psd) {PSD Est.};
	
	\node[loudspeaker style,rotate=-90] (speaker) at (3.75,-.5) {};
	\node[microphone style,rotate=90] (mic) at (3.75,-2.25) {};
	
	\draw [thick] ($(sigInX1)$) -| (speaker.west) node [below, pos=-.119] {\commentTHa{$\underline{\boldsymbol{x}}_{\tau}$}};
	%\draw [->, thick] (sigInX) -| (pbkf);
	\draw [->, thick] (pbkf) -- (subtraction.north) node [pos=.4, right] {$\widehat{\underline{\boldsymbol{d}}}_{\text{early},\tau}$} node [pos=.8, right] {$-$};
	\draw [->, thick] (mic.west) |- (subtraction.east) node [pos=.85, above] {$\underline{\boldsymbol{y}}_{\tau}$};
	\draw [->, thick] ($(pbkf.north)+(0,.74)$) -- (pbkf.north);
	
	\draw [->, thick] (subtraction.west) -- ($(pf.east)+(0,-.1)$) node [pos=.3, above] {$\underline{\boldsymbol{e}}_{\tau}^+$};
	
	\draw [->, thick] (pf.west) -- (epf) node [pos=.5, above]  {$\hat{\underline{\boldsymbol{s}}}_{\tau}$};
	
	%\draw [-, thick] (sigInX1) -| (pbkf);
	\draw [->, thick] ($(sigInX1)+(0.2,.0)$) |- ($(pf.east) + (0,.1)$);
	
	\draw [->, thick] ($(pf.north)-(-0,0)$) |- (psd.west) node [pos=.3, left] {$\widehat{\boldsymbol{M}}_\tau$};
	\draw [thick, <-] (psd.south) -- ($(psd.south)+(0,-1.65)$);
	\draw [->, thick] (psd.east) -- (pbkf.west) node [pos=.5, above] {$\commentTHg{\hat{\boldsymbol{\Psi}}_{\tau}^{\text{I}}}$};

	\draw [thick] (2.625,-2.7) rectangle (4.5,-.25);

	% \draw [thick] (subtraction.east) -| (mic.west);
	% \draw [thick] (sigInX1) -| (speaker.west);
	
	\draw [thick] ($(sigInX1)+(-1.8,0)$) -- ($(sigInX1)$);
	
	\draw [thick, dashed, ->] ($(speaker)-(.0,.2)$) -- ($(mic)+(.0,.2)$);
	\draw [thick, dashed, ->] ($(speaker)-(.0,.2)$) -- ($(speaker)+(.75,-.8)$) -- ($(mic)+(.1,.2)$);
	
	\node [] (d) at ($(mic)+(.55,.3)$) {$\underline{\boldsymbol{d}}_\tau$}; 
	
	% \node [] (s) at ($(mic)+(-.55,.3)$) {$\boldsymbol{d}_\tau$};
	\draw [fill] (3.22,-1.12) circle (.05);
	\draw [thick] (3.1,-1.2) arc(225:360:.15);
	\node [] (s) at ($(3.22,-1.70)$) {$\underline{\boldsymbol{s}}_\tau$};
	\draw [dashed, ->, thick] ($(3.3,-1.3)+(0,0)$) -- ($(mic)+(-.1,.2)$);

	\draw [fill] (2.8,-2.2) circle (.05);
	\draw [thick] (2.9,-2.3) arc(-45:45:.15);
	\draw [dashed, ->, thick] ($(3.0,-2.2)+(0,0)$) -- ($(mic)+(-.2,.10)$);
	\node [] (s) at ($(3.32,-2.45)$) {$\underline{\boldsymbol{n}}_\tau$};
	
	\end{tikzpicture}
	\caption{Proposed synergistic \ac{KF}+\ac{DPF} approach to \ac{AEC}.}	
	\label{fig:alg_overview}
\end{figure}
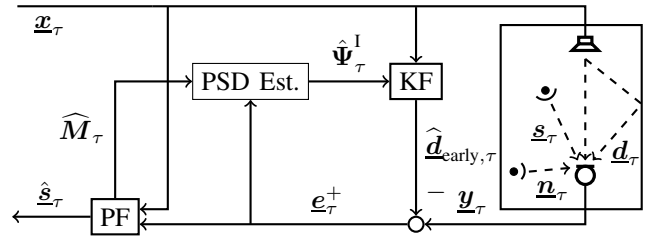

In the following, we use bold uppercase letters for matrices and bold lowercase letters for vectors with underlined symbols indicating time-domain quantities. A matrix element in the $m$th row and the $n$th column is indicated by $[\cdot]_{mn}$. \commentTHa{We} denote the all-zero matrix of dimensions $D_1 \times D_2$ by $\boldsymbol{0}_{D_1 \times D_2}$, the $D$-dimensional identity matrix by $\boldsymbol{I}_D$ and the $D$-dimensional \ac{DFT} matrix by $\boldsymbol{F}_D$. Furthermore, the transposition and Hermitian transposition are represented by $(\cdot)^{\text{T}}$ and $(\cdot)^{\text{H}}$, respectively. The proper complex Gaussian \ac{PDF}, with mean vector $\boldsymbol{\mu}$ and covariance matrix $\boldsymbol{\Psi}$, is denoted by $\mathcal{N}_c(\cdot| \boldsymbol{\mu}, \boldsymbol{\Psi})$ and the expectation operator by $\mathbb{E}[\cdot]$. Finally, the $\text{diag}(\cdot)$ operator constructs a diagonal matrix from its vector-valued argument. %\shorten{the mathematical part is quite long here!}

\section{Probabilistic Signal Model}
\label{sec:prob_sig_mod}
%
%\begin{itemize}
%	\item block-based time-domain signal model
%	\item describe early echo, late echo, background noise and near-end interferer (maybe refer already to image)
%	\item describe model by time-varying Gaussian
%	\item discuss the PSD properties in terms of stationarity
%	\item discuss the random walk filter model
%	\item describe state-space model
%\end{itemize}
%
%\vspace*{.5cm}
%
The observed time-domain microphone signal block ${\underline{\boldsymbol{y}}}_\tau$ is modelled as a linear superposition of an early echo component $\underline{\boldsymbol{d}}_{\text{early},\tau}$, a late echo component  $\underline{\boldsymbol{d}}_{\text{late},\tau}$, background noise $\underline{\boldsymbol{n}}_{\tau}$ and a near-end speaker $\underline{\bs}_\tau$ (cf.~Fig.~\ref{fig:alg_overview}), as follows:
\begin{equation}
	{\underline{\boldsymbol{y}}}_\tau = \underline{\boldsymbol{d}}_{\text{early},\tau} + \underline{\boldsymbol{d}}_{\text{late},\tau} +  \underline{\boldsymbol{n}}_{\tau} + \underline{\bs}_\tau \in \mathbb{R}^{R} 
	\label{eq:td_sig_mod}
\end{equation}
%
%at block index $\tau$ is modelled as a linear superposition of an early echo component $\underline{\boldsymbol{d}}_{\text{early},\tau}$, a late echo component  $\underline{\boldsymbol{d}}_{\text{late},\tau}$, background noise $\underline{\boldsymbol{n}}_{\tau}$ and a near-end speaker $\underline{\bs}_\tau$ (cf.~Fig.~\ref{fig:alg_overview}). Each signal block ${\underline{\boldsymbol{y}}}_\tau$ in Eq.~\eqref{eq:td_sig_mod} is composed of $R$ consecutive samples, i.e., 
where signal block ${\underline{\boldsymbol{y}}}_\tau$ at time index $\tau$ consists of $R$ consecutive samples, i.e.,
\begin{equation}
	\underline{\boldsymbol{y}}_\tau = \begin{bmatrix}
		\underline{y}_{\tau R - R + 1}, \underline{y}_{\tau R - R + 2}, \dots, \underline{y}_{\tau R}
	\end{bmatrix}^{{\text{T}}} \in \mathbb{R}^{R},
\end{equation}
and $\underline{\boldsymbol{d}}_{\text{early},\tau}$, $\underline{\boldsymbol{d}}_{\text{late},\tau}$, $\underline{\bs}_\tau$ and $\underline{\boldsymbol{n}}_{\tau}$ are defined analogously. The early echo component ${\underline{\boldsymbol{d}}}_{\text{early}, \tau}$ is modelled by a linear convolution of a \ac{FIR} filter $\underline{\boldsymbol{w}}_{\tau} \in \mathbb{R}^{L}$ with the corresponding far-end signal block $\underline{\boldsymbol{x}}_{\tau}\in \mathbb{R}^{L+R-1}$. The linear convolution can efficiently be implemented by a \ac{PBC}. For this the \ac{FIR} filter $\underline{\boldsymbol{w}}_{\tau}$ is separated into $B=\frac{L}{R}$ partitions $\underline{\w}_{b,\tau} \in \mathbb{R}^R$. Subsequently, each partition is convolved with a corresponding delayed far-end block
\begin{equation}
	\underline{\boldsymbol{x}}_{b,\tau} = \begin{bmatrix}
		\underline{x}_{(\tau-b) R - M + 1}, \underline{x}_{(\tau-b) R - M + 2}, \dots, \underline{x}_{(\tau-b) R}
	\end{bmatrix}^{{\text{T}}} \in \mathbb{R}^{M} 
	\label{eq:x_def}
\end{equation}
of length $M=2R$ and the convolution products are added. By implementing each linear convolution in the \ac{DFT} domain one obtains \commentTHa{\cite{kuech_state-space_2014}}:
\begin{equation}
	{\underline{\boldsymbol{d}}}_{\text{early}, \tau} = \sum_{b=0}^{B-1} \boldsymbol{Q}_1^{{\text{T}}} \boldsymbol{F}_M^{-1} \boldsymbol{X}_{b,\tau} {\w}_{b,\tau} \in \mathbb{R}^{R} 
	% {\underline{\boldsymbol{d}}}_{\text{early}, \tau} = \sum_{b=0}^{B-1} \boldsymbol{Q}_1^{{\text{T}}} \boldsymbol{F}_M^{-1} \boldsymbol{X}_\tau \boldsymbol{F}_M \boldsymbol{Q}_2 \underline{\w}_\tau \in \mathbb{R}^{R} 
	\label{eq:timeDomObsEq}
\end{equation}
with the constraint matrix \makebox{$\boldsymbol{Q}_1^{\text{T}} = \begin{bmatrix}\boldsymbol{0}_{R \times R} & \boldsymbol{I}_R\end{bmatrix}$}, the \ac{DFT}-domain \ac{FIR} filter partition ${\w}_{b,\tau} = \boldsymbol{F}_M \boldsymbol{Q}_2 \underline{\w}_{b,\tau} \in \mathbb{C}^M $, the \ac{DFT}-domain far-end signal block \makebox{$\X_{b,\tau} = \text{diag} \left( \boldsymbol{F}_M \underline{\boldsymbol{x}}_{b,\tau} \right)$} and the zero-padding matrix \makebox{$\boldsymbol{Q}_2^{\text{T}}= \begin{bmatrix}\boldsymbol{I}_{R} & \boldsymbol{0}_{R \times R }\end{bmatrix}$}.
%	
%
%\vspace*{1cm}
%
%by a linear \ac{PBC} \cite{Stockham1966HighspeedCA} of $B$ far-end signal blocks 
%%
%\begin{equation}
%\underline{\boldsymbol{x}}_{\tau-b} = \begin{bmatrix}
%\underline{x}_{(\tau-b) R - M + 1}, \underline{x}_{(\tau-b) R - M + 2}, \dots, \underline{x}_{(\tau-b) R}
%\end{bmatrix}^{{\text{T}}} \in \mathbb{R}^{M} 
%\label{eq:x_def}
%\end{equation}
%%
%of even length $M=2R$ with $B$ time-domain \ac{FIR} filter partitions $\underline{\w}_{b,\tau} \in \mathbb{R}^{R}$. The \ac{PBC} in Eq.~\eqref{eq:timeDomObsEq} is described by an overlap-save convolution using the \ac{DFT} {domain} far-end signal blocks \makebox{$\X_{\tau-b} = \text{diag} \left( \boldsymbol{F}_M \underline{\boldsymbol{x}}_{\tau-b} \right)$}, the zero-padding matrix \makebox{$\boldsymbol{Q}_2^{\text{T}}= \begin{bmatrix}\boldsymbol{I}_{M-R} & \boldsymbol{0}_{M-R \times R }\end{bmatrix}$} and the linear convolution constraint matrix \makebox{$\boldsymbol{Q}_1^{\text{T}} = \begin{bmatrix}\boldsymbol{0}_{R \times M-R} & \boldsymbol{I}_R\end{bmatrix}$}. 
% By first inserting the \ac{PBC} model \eqref{eq:timeDomObsEq} into the signal model \eqref{eq:td_sig_mod} and afterwards transforming the zero-padded time-domain signals into the \ac{DFT} domain, i.e., ${\boldsymbol{y}}_\tau = \boldsymbol{F}_M \boldsymbol{Q}_1 \underline{\boldsymbol{y}}_\tau$, one obtains the frequency-domain observation equation % TODO: similar formulation as in ICASSP paper
%
A relation between the noisy observation $\underline{\boldsymbol{y}}_\tau$ and the filter partitions $\boldsymbol{w}_{b, \tau}$ is obtained by inserting the \ac{PBC} model \eqref{eq:timeDomObsEq} into the signal model \eqref{eq:td_sig_mod}. By using ${\boldsymbol{y}}_\tau = \boldsymbol{F}_M \boldsymbol{Q}_1 \underline{\boldsymbol{y}}_\tau$ this time-domain observation equation \commentTHa{is transformed} to the \ac{DFT} domain:
\begin{equation}
	{{\boldsymbol{y}}}_\tau =  \sum_{b=0}^{B-1} \boldsymbol{C}_{b,\tau} {\w}_{b,\tau}  + {\boldsymbol{d}}_{\text{late},\tau} +  {\boldsymbol{n}}_{\tau} + {\bs}_\tau \in \mathbb{C}^{M} 
	\label{eq:fd_sig_mod}
\end{equation}
with the \ac{DFT}-domain signal components $ {\boldsymbol{d}}_{\text{late},\tau}$, ${\boldsymbol{n}}_{\tau}$, and $ {\bs}_\tau$ and the overlap-save-constrained far-end signal blocks \makebox{$\boldsymbol{C}_{b,\tau} = \boldsymbol{F}_M \boldsymbol{Q}_1 \boldsymbol{Q}_1^{\text{T}} \boldsymbol{F}_M^{-1} \X_{b,\tau}$}. Note that the corresponding time-domain signals can be computed by the inverse transform \makebox{${\underline{\boldsymbol{y}}}_{\tau} = \boldsymbol{Q}_1^{\text{T}}\boldsymbol{F}_M^{-1}{\boldsymbol{y}}_{\tau}$}. The late echo component ${\boldsymbol{d}}_{\text{late},\tau}$, the background noise ${\boldsymbol{n}}_{\tau}$, and the near-end speaker ${\bs}_\tau$ are considered as additive disturbances when estimating the \ac{AF} partitions ${\w}_{b,\tau}$ in Eq.~\eqref{eq:fd_sig_mod}. In the following, we model each of these interfering components as a zero-mean, non-stationary and spectrally uncorrelated proper complex Gaussian random process by the respective \acp{PDF}%. %{The respective signal blocks are distributed according to} %the \acp{PDF}  % TODO: similar formulation as in ICASSP paper
\begin{align} % TODO: write as one equation
	p({\boldsymbol{d}}_{\text{late}, \tau})  &= \mathcal{N}_c({\boldsymbol{d}}_{\text{late},\tau}|\boldsymbol{0}_{M \times 1}, {{\boldsymbol{\Psi}}}_{\tau}^{\text{D}_{\text{late}}}) \label{eq:d_late_psd} \\
	p({\boldsymbol{n}}_\tau)  &= \mathcal{N}_c({\boldsymbol{n}}_{\tau}|\boldsymbol{0}_{M \times 1}, {\boldsymbol{\Psi}}_{\tau}^{\text{N}}), \label{eq:bg_psd}\\
	p({\bs}_\tau)  &= \mathcal{N}_c({\bs}_{\tau}|\boldsymbol{0}_{M \times 1}, {\boldsymbol{\Psi}}_{\tau}^{\text{S}}), \label{eq:nearend_psd}
\end{align}
%
%$p({\bs}_\tau)  = \mathcal{N}_c({\bs}_{\tau}|\boldsymbol{0}_{M \times 1}, {\boldsymbol{\Psi}}_{\tau}^S)$, $p({\boldsymbol{d}}_{\text{late}, \tau})= \mathcal{N}_c({\boldsymbol{d}}_{\text{late},\tau}|\boldsymbol{0}_{M \times 1}, {\boldsymbol{\Psi}}_{\tau}^{D_{\text{late}}})$ and $p({\boldsymbol{n}}_\tau)= \mathcal{N}_c({\boldsymbol{n}}_{\tau}|\boldsymbol{0}_{M \times 1}, {\boldsymbol{\Psi}}_{\tau}^N)$ 
with the diagonal \ac{PSD} matrices ${\boldsymbol{\Psi}}_{\tau}^{\text{D}_{\text{late}}}$,  ${\boldsymbol{\Psi}}_{\tau}^{\text{N}}$ and \makebox{${\boldsymbol{\Psi}}_{\tau}^{\text{S}} \in \mathbb{C}^{M \times M}$}. \commentTHa{We} assume ${\boldsymbol{d}}_{\text{late}, \tau}$, ${\boldsymbol{n}}_\tau$ and ${\boldsymbol{\bs}}_\tau$  to be mutually uncorrelated
\begin{equation}
	\mathbb{E}\left[{\boldsymbol{d}}_{\text{late},\tau} {\boldsymbol{n}}_{\tau}^{\text{H}} \right]= \mathbb{E}\left[{\boldsymbol{d}}_{\text{late},\tau} {\boldsymbol{s}}_{\tau}^{\text{H}} \right]= \mathbb{E}\left[{\boldsymbol{s}}_{\tau} {\boldsymbol{n}}_{\tau}^{\text{H}} \right] = \commentTHf{\boldsymbol{0}_{M \times M}}.
\end{equation}
%.
%, i.e., $\mathbb{E}\left[{\boldsymbol{d}}_{\text{late},\tau} {\boldsymbol{n}}_{\tau}^{\text{H}} \right]= \mathbb{E}\left[{\boldsymbol{d}}_{\text{late},\tau} {\boldsymbol{s}}_{\tau}^{\text{H}} \right]= \mathbb{E}\left[{\boldsymbol{s}}_{\tau} {\boldsymbol{n}}_{\tau}^{\text{H}} \right] =\boldsymbol{0}_{R \times R}$.
%
Finally, to account for the time-variance of acoustic environments, the temporal evolution of each \ac{AF} partition ${\w}_{b,\tau}$ is modelled by a \ac{DFT}-domain random walk Markov model \cite{kuech_state-space_2014}
\begin{align}
	{\w}_{b,\tau} &= A  ~{\w}_{b,\tau - 1}  + \Delta {{\w}}_{b,\tau} \label{eq:stateTransMod}
	% &\text{with}& \quad  \Delta {{\w}}_{b,\tau} \sim   \mathcal{N}_c(\Delta {{\w}}_{b,\tau}|{\boldsymbol{0}_{M \times 1}}, \boldsymbol{\Psi}_{b,\tau}^\Delta)
	%{\y}_{\tau} &= \sum_{b=0}^{B-1}{\C}_{\tau-b}  {\w}_{b,\tau} + {\tilde{\boldsymbol{n}}}_{\tau} &\text{with} & ~~~\quad {\tilde{\boldsymbol{n}}}_{\tau} \sim   \mathcal{N}_c({\bs}_{\tau}|{\boldsymbol{0}_{M \times 1}}, \boldsymbol{\Psi}_{\tau}^{I), \label{eq:obsMod}
\end{align}
with the process noise vector of the $b$th partition $\Delta {{\w}}_{b,\tau}$ and the state transition coefficient $0 < A < 1$. The process noise $\Delta {{\w}}_{b,\tau}$ is assumed to be distributed according to 
\begin{equation}
	p(\Delta {{\w}}_{b,\tau}) = \mathcal{N}_c(\Delta {{\w}}_{b,\tau}|{\boldsymbol{0}_{M \times 1}}, {\boldsymbol{\Psi}}_{b,\tau}^{\Delta\text{W}}) \label{eq:process_noise_model}
\end{equation}
%
%\makebox{$p(\Delta {{\w}}_{b,\tau}) = \mathcal{N}_c(\Delta {{\w}}_{b,\tau}|{\boldsymbol{0}_{M \times 1}}, {\boldsymbol{\Psi}}_{b,\tau}^W)$} 
with the diagonal process noise \commentTHb{\ac{PSD}} matrix ${\boldsymbol{\Psi}}_{b,\tau}^{\Delta\text{W}}$. Note that \commentTHf{Eqs.~\eqref{eq:fd_sig_mod} - \eqref{eq:process_noise_model}} represent a linear Gaussian state-space model with the \ac{DFT}-domain \ac{AF} partitions ${\w}_{b,\tau}$ as the states and the microphone signal blocks $\boldsymbol{y}_\tau$ as the observations.

\section{Acoustic Echo Cancellation Model}
\label{sec:aec_alg}
% TODO: evtl pfeifenberger besser referenzieren
The considered \ac{AEC} architecture is depicted in Fig.~\ref{fig:alg_overview}. A linear \ac{AF} estimates the early echo component $\underline{\boldsymbol{d}}_{\text{early}, \tau}$ which is then subtracted from the noisy observation $\underline{\boldsymbol{y}}_\tau$. The estimation error signal $\underline{\boldsymbol{e}}_\tau^+$ is used as input to a \ac{DPF}. Finally, for a double-talk robust adaptation control of the \ac{AF} partitions $\boldsymbol{w}_{b,\tau}$, we propose a novel approach to estimate the observation noise \ac{PSD} matrix by exploiting the near-end speaker estimate of the \ac{DPF} and the estimation error $\underline{\boldsymbol{e}}_\tau^+$.

\subsection{Adaptive Kalman Filter}
\label{sec:pbkf}
%
%\begin{itemize}
%	\item we model the adaptive filter partition to be proper complex Gaussian distributed  with diagonal state uncertainty
%	\item each partition is modelled to be uncorrelated to each other
%	\item due to the Gaussian model of the adaptive filter coefficients and the noise model closed-form inference by a Kalman filter equations is possible
%	\item Introduce Kalman filter equations
%	\item discuss the diagonal step size and what it does
%	\item noise-robustness and reconvergence properties highly depend on a precise estimation of the observation noise PSD
%	\item this is usually done by a recursive aveage of the prior error (only diagonal elements)
%	\item problem that for rapidly changing acoustic environments the observation noise is largely overestimated which results in slow reconvergence
%\end{itemize}
%
% TODO: alternatively you could not use we model: but just state the respective paper and the equations?
%
% TODO \textcolor{red}{to be more innovative you could write in terms of your model (which is different in comparison the model of kuech and afterwards mentiond that by summing up the various PSDs you get the same?)}
%
We model the state posterior of each \ac{AF} partition ${\boldsymbol{w}}_{b,\tau}$, given all preceding observations ${\boldsymbol{Y}}_{1:\tau} = \begin{bmatrix}{\boldsymbol{y}}_{1}, & \dots, & {\boldsymbol{y}}_{\tau}\end{bmatrix}$, by
\begin{equation}
	p({\boldsymbol{w}}_{b,\tau} | {\boldsymbol{Y}}_{1:\tau} ) = \mathcal{N}_c \left({\boldsymbol{w}}_{b,\tau}|\hat{\boldsymbol{w}}_{b,\tau}, {\bP}_{b,\tau}   \right)
\end{equation}
%
%\begin{equation}
%p({\boldsymbol{w}}_{\tau} | {\boldsymbol{Y}}_{1:\tau} ) = \prod_{b=0}^{B-1} p({\boldsymbol{w}}_{b,\tau} | {\boldsymbol{Y}}_{1:\tau} )= \prod_{b=0}^{B-1} \mathcal{N}_c \left({\boldsymbol{w}}_{b,\tau}|\hat{\boldsymbol{w}}_{b,\tau}, {\bP}_{b,\tau}   \right)
%\end{equation}
%{We model the state posterior of the \ac{AF} partitions ${\boldsymbol{w}}_{b,\tau}$, with $b=0,\dots,B-1$, given all preceding observations ${\boldsymbol{Y}}_{1:\tau} = \begin{bmatrix}{\boldsymbol{y}}_{1}, & \dots, & {\boldsymbol{y}}_{\tau}\end{bmatrix}$, by \cite{kuech_state-space_2014}}
%%
%\begin{align}
%p({\boldsymbol{w}}_{0,\tau}, \dots, {\boldsymbol{w}}_{B-1,\tau}  | {\boldsymbol{Y}}_{1:\tau} ) &= \prod_{b=0}^{B-1} p({\boldsymbol{w}}_{b,\tau} | {\boldsymbol{Y}}_{1:\tau} ) \\&= \mathcal{N}_c \left({\boldsymbol{w}}_{b,\tau}|\hat{\boldsymbol{w}}_{b,\tau}, {\bP}_{b,\tau}   \right)
%\end{align}
% TODO: check this statement
with mean $\hat{\boldsymbol{w}}_{b,\tau}$ and diagonal state uncertainty matrix ${\bP}_{b,\tau}$. Due to the linear Gaussian model (cf. Eqs. \eqref{eq:fd_sig_mod} - \eqref{eq:process_noise_model}), a closed-form inference of the state posterior is given by the \ac{KF} equations. By setting the transition factor to one in the prediction of the \commentTHf{echo} and introducing a gradient constraint, the diagonalized \ac{PBKF} is obtained \cite{kuech_state-space_2014}
%
%\textcolor{red}{Maybe write something like inference leads to Kalman filter that can be interpreted as partitioned block adaptive filter with adaptive step size. check if kuech uses A in prediction? check if the results change?}
% TODO: ch
%
\begin{align} % TODO: check this again and maybe take A out in prediction
	\label{eq:eStep}
	%\hat{\boldsymbol{w}}^{+}_{b,\tau - 1} &= \hat{\boldsymbol{w}}_{b,\tau - 1} \notag \\ % maybe take A out here
	&	\widehat{\boldsymbol{d}}_{\text{early},\tau}=\sum_{b=0}^{B-1} {\C}_{b,\tau} \hat{\boldsymbol{w}}_{b,\tau - 1}  \commentTHf{\approx A~\sum_{b=0}^{B-1} {\C}_{b,\tau} \hat{\boldsymbol{w}}_{b,\tau - 1}} \notag \\
	&	{\e}_{\tau}^{+} 					=  {\y}_{\tau} - \widehat{\boldsymbol{d}}_{\text{early},\tau} \notag\\
	&{\bP}^{+}_{b,\tau - 1} 				= A^2 ~ {\bP}_{b,\tau - 1} + {\boldsymbol{\Psi}}^{\Delta\text{W}}_{b,\tau} \\
	% &\boldsymbol{\Lambda}_{b,\tau} 		= {\bP}^{+}_{b,\tau - 1} \left(\sum_{b=0}^{B-1} {\X}_{\tau-b} {\bP}^{+}_{b,\tau-1} {\X}_{\tau-b}^{\herm}  + \frac{M}{R} {\boldsymbol{\Psi}}^{\text{I}}_{\tau}\right)^{-1} \notag\\
		&\boldsymbol{\Lambda}_{b,\tau} 		= {\bP}^{+}_{b,\tau - 1} \left(\commentTHe{\sum_{\tilde{b}=0}^{B-1} {\X}_{\tilde{b},\tau} {\bP}^{+}_{\tilde{b},\tau-1} {\X}_{\tilde{b},\tau}^{\herm}} + \frac{M}{R} {\boldsymbol{\Psi}}^{\text{I}}_{\tau}\right)^{-1} \notag\\
	% \hat{\boldsymbol{w}}_{b,\tau}	 	& = \hat{\boldsymbol{w}}^{+}_{b,\tau - 1} + \boldsymbol{G} \boldsymbol{\Lambda}_{b,\tau} {\X}_{b,\tau}^{\herm} {\e}_{\tau}^{+} \notag\\
	&\hat{\boldsymbol{w}}_{b,\tau}	 	 = \hat{\boldsymbol{w}}_{b,\tau - 1} + \boldsymbol{G} \boldsymbol{\Lambda}_{b,\tau} {\X}_{b,\tau}^{\herm} {\e}_{\tau}^{+} \notag\\
	&{\bP}_{b,\tau} 						 =  \left({\I}_M - \frac{R}{M} \boldsymbol{\Lambda}_{b,\tau} {\X}_{b,\tau}^{\herm} {\X}_{b,\tau}\right) {\bP}^{+}_{b,\tau-1} \notag
\end{align}
with the estimated early echo signal $\widehat{\boldsymbol{d}}_{\text{early},\tau}$, the prior error ${\e}_{\tau}^{+}$, the \commentTHa{gradient} constraint matrix $\boldsymbol{G} = \boldsymbol{F}_M \boldsymbol{Q}_2 \boldsymbol{Q}_2^{\text{T}} \boldsymbol{F}_M^{-1}$ and the \commentTHa{adaptive diagonal step size} matrix $\boldsymbol{\Lambda}_{b,\tau}$. The double-talk robustness and convergence properties of the \ac{KF} crucially depend on \commentTHa{a precise estimation of the} observation noise \ac{PSD} matrix \makebox{${\boldsymbol{\Psi}}^{\text{I}}_{\tau} = {\boldsymbol{\Psi}}^{\text{D}_{\text{late}}}_{\tau} + {\boldsymbol{\Psi}}^{\text{N}}_{\tau} + {\boldsymbol{\Psi}}^{S}_{\tau}$} \commentTHa{(cf. Sec.~\ref{sec:psd_est}).} % \commentTHa{from the observed data (see Sec.~\ref{sec:psd_est}).}

\subsection{Deep Neural Network-based Postfilter}
\label{sec:deep_pf}
%The aim of any \ac{PF} is the estimation of the near-end signal $\underline{\boldsymbol{s}}_\tau$ \commentTHg{(cf.~Eq.~\eqref{eq:td_sig_mod})} given the estimated error \commentTHg{\makebox{$\underline{\boldsymbol{e}}_{\tau}^+=\boldsymbol{Q}_1^{\text{T}} \boldsymbol{F}_M^{-1} {\boldsymbol{e}}_{\tau}^+$}} \commentTHg{(cf.~Eq.~\eqref{eq:eStep})}. 
% The aim of any \ac{PF} is the \commentTHa{estimation of the near-end signal $\underline{\boldsymbol{s}}_\tau$}. 
The aim of any \ac{PF} is the estimation of the near-end signal $\underline{\boldsymbol{s}}_\tau$ \commentTHg{(cf.~Eq.~\eqref{eq:td_sig_mod})} given the estimated error \commentTHg{\makebox{$\underline{\boldsymbol{e}}_{\tau}^+=\boldsymbol{Q}_1^{\text{T}} \boldsymbol{F}_M^{-1} {\boldsymbol{e}}_{\tau}^+$}} \commentTHg{(cf.~Eq.~\eqref{eq:eStep})}. 
\commentTHa{We} consider a recurrent \commentTHb{\ac{DNN}}-based \ac{PF} which is inspired by \cite{pfeifenberger_nonlinear_2020, Xia2020} and comprises four layers. The first layer is a dense feedforward layer with tanh activations which combines the input features to a vector of dimension $P$. Subsequently, two stacked \commentTHg{\ac{GRU}} layers are added which extract temporal information from the combined features. Finally, a dense feedforward output layer with sigmoid activations is used to map the \commentTHg{\ac{GRU}} states to a corresponding frame-wise diagonal masking matrix $\widehat{\boldsymbol{M}}_\tau$. % The vector containing all \commentTHb{\ac{DNN}} parameters is denoted by $\boldsymbol{\theta}$. 

% Regarding the input features $\tilde{\boldsymbol{u}}_\tau$ to the \commentTHb{\ac{DNN}}, we use the normalized logarithmic power spectra of the prior error and the far-end signal. In particular, we first compute the windowed \ac{STFT} of the time-domain far-end signal and the prior error (cf.~Sec.~\ref{sec:prob_sig_mod}),
\commentTHa{As} input features $\tilde{\boldsymbol{u}}_{\text{feat},\tau}$ to the \commentTHb{\ac{DNN}}, we use the normalized logarithmic power spectra of the prior error and the far-end signal. \commentTHa{For this}, we first compute the \commentTHa{\ac{STFT}} of the \commentTHa{time-domain signals} (cf.~Sec.~\ref{sec:prob_sig_mod}),
\begin{align}
	\tilde{\boldsymbol{u}}_{\tau} = \begin{bmatrix}
		\tilde{\boldsymbol{e}}_{\tau}^+ \\
		\tilde{\boldsymbol{x}}_{0,\tau}
	\end{bmatrix} = \begin{bmatrix}
		\boldsymbol{F}_M \boldsymbol{V} \begin{bmatrix} (\underline{\boldsymbol{e}}_{\tau-1}^+)^{\text{T}} & (\underline{\boldsymbol{e}}_{\tau}^+)^{\text{T}} \end{bmatrix}^{\text{T}} \\
		\boldsymbol{F}_M \boldsymbol{V} \underline{\boldsymbol{x}}_{0,\tau}
	\end{bmatrix} \in \mathbb{C}^{2M}
	\label{eq:feature_e_x}
\end{align}
with the diagonal window matrix $\boldsymbol{V} \in \mathbb{R}^{M \times M}$ and $\tilde{(\cdot)}$ denoting windowed \ac{STFT}-domain quantities. Subsequently, the normalized logarithmic power spectrum is computed by
\begin{equation}
	%[\tilde{\boldsymbol{u}}_{\text{feat},\tau}]_m = \frac{\text{max}(\log (| [\tilde{\boldsymbol{u}}_{,\tau}]_m|^{2}), \epsilon) - [\boldsymbol{\mu}]_{m}}{ [\boldsymbol{\sigma}]_m},
		[\tilde{\boldsymbol{u}}_{\text{feat},\tau}]_m = \frac{\commentTHa{\log (\text{max}(| [\tilde{\boldsymbol{u}}_{\tau}]_m|^{2}, \epsilon_1))} - [\boldsymbol{\mu}]_{m}}{ [\boldsymbol{\sigma}]_m},
\end{equation}
with $m=0,\dots,2M-1$ and $\commentTHa{\epsilon_1}>0$ being a small number to avoid numerical instabilities. Here, the estimated mean and standard deviation of the feature vector are denoted by $\boldsymbol{\mu}$ and $\boldsymbol{\sigma}$, respectively. Note that due to the symmetry of the \ac{STFT} only the non-redundant \commentTHg{$M+2$} frequency bins of $\tilde{\boldsymbol{u}}_{\text{feat},\tau}$ are used as features.
%
% The \commentTHb{\ac{DNN}} is trained by minimizing the cost function \cite{carbajal_joint_2020, nugraha_multichannel_2016}
%
\commentTHa{The parameters $\boldsymbol{\theta}$ of the \commentTHb{\ac{DNN}} are} trained by minimizing the cost function \cite{carbajal_joint_2020, nugraha_multichannel_2016}
\begin{equation}
	\mathcal{J}_{\text{PF}}(\boldsymbol{\theta}) = \sum_{\tau,m} d_{\text{KL}}(|\tilde{s}_{m \tau}| ,  |\hat{\tilde{s}}_{m \tau}| ),
	\label{eq:cost_func}
\end{equation}
defined by the \acl{KL} divergence
\begin{equation}
	d_{\text{KL}}(|\tilde{s}_{m \tau}| ,  |\hat{\tilde{s}}_{m \tau}| ) = - |\tilde{s}_{m \tau}|   \log(|\hat{\tilde{s}}_{m \tau}|  + \commentTHa{\epsilon_2}) + |\hat{\tilde{s}}_{m \tau}| 
	\label{eq:kl_div}
\end{equation}
between the magnitudes of \commentTHa{the} true \ac{STFT} near-end component \makebox{$ \tilde{s}_{m\tau} = \left[ \boldsymbol{F}_M \boldsymbol{V} \begin{bmatrix} (\underline{\boldsymbol{s}}_{\tau-1})^{\text{T}} & (\underline{\boldsymbol{s}}_{\tau})^{\text{T}} \end{bmatrix}^{\text{T}}\right]_m $} and the estimated one \makebox{$\hat{\tilde{{s}}}_{m \tau} = [\widehat{\boldsymbol{M}}_{\tau}]_{mm} [\tilde{\boldsymbol{e}}_{\tau}^+]_m$}.
All constant terms have been omitted in Eq.~\eqref{eq:kl_div} and a regularization term \commentTHa{$\epsilon_2$} has been \commentTHa{included} \cite{nugraha_multichannel_2016}. 
\commentTHg{Note that any mask-based \ac{PF} could be used to support the observation noise \ac{PSD} estimation of the \ac{KF} ({cf.}~Sec.~\ref{sec:psd_est}).}

\section{Proposed Power Spectral Density Estimation}
\label{sec:psd_est}
%
% Decisive for a fast-converging and double-talk robust adaptation of the adaptive filter partitions $\hat{\boldsymbol{w}}_{b,\tau}$ is a precise estimation of the the process noise covariance matrix $\boldsymbol{\Psi}_{b,\tau}^{W}$ and the observation noise covariance matrix $\boldsymbol{\Psi}_{\tau}^{I}$.
% For a fast-converging and double-talk robust adaptation of the \ac{AF} partitions $\hat{\boldsymbol{w}}_{b,\tau}$, a precise estimation of the observation noise \ac{PSD} matrix ${\boldsymbol{\Psi}}^{\text{I}}_{\tau}$ and the partition-wise process noise \ac{PSD} matrix ${\boldsymbol{\Psi}}_{b,\tau}^{\Delta\text{W}}$ is decisive. 
For a fast-converging and double-talk robust adaptation of the \ac{AF} partitions $\hat{\boldsymbol{w}}_{b,\tau}$, a precise estimation of the observation noise \ac{PSD} matrix ${\boldsymbol{\Psi}}^{\text{I}}_{\tau}$ and \commentTHa{process noise \ac{PSD} matrices} ${\boldsymbol{\Psi}}_{b,\tau}^{\Delta\text{W}}$ is decisive. In particular, the observation noise \ac{PSD} matrix ${\boldsymbol{\Psi}}^{\text{I}}_{\tau}$ is, due to its fast changing behaviour, difficult to estimate. In contrast to state-of-the-art approaches, we suggest to exploit the different statistical properties of the signal components generating the observation $\boldsymbol{y}_\tau$ (cf.~Eqs.~\eqref{eq:td_sig_mod}~and~\eqref{eq:fd_sig_mod}).

We start by representing the prior error signal
\begin{equation}
	\boldsymbol{e}_\tau^+ = {\boldsymbol{e}}_{\text{early},\tau}^+ + \boldsymbol{p}_{\tau} +  \boldsymbol{s}_{\tau} 
\end{equation}
in terms of the early echo error signal \makebox{$ {\boldsymbol{e}}_{\text{early},\tau}^+   = {\boldsymbol{d}}_{\text{early},\tau} - \widehat{\boldsymbol{d}}_{\text{early},\tau}$}, the late reverberation and background noise signal \makebox{$\boldsymbol{p}_{\tau}=\boldsymbol{d}_{\text{late},\tau}+\boldsymbol{n}_{\tau}$} and the desired near-end speaker signal $\boldsymbol{s}_{\tau}$. By assuming the early echo error ${\boldsymbol{e}}_{\text{early},\tau}^+ $, the noise signal $\boldsymbol{p}_{\tau}$ and the near-end speaker signal $\boldsymbol{s}_{\tau} $ to be mutually uncorrelated, the \ac{PSD} matrix of the prior error $\boldsymbol{e}^+_\tau$ is given by
\begin{equation}
	{\boldsymbol{\Psi}}_{\tau}^{\text{E}} = {\boldsymbol{\Psi}}_{\tau}^{{\text{E}}_{\text{early}}} + \boldsymbol{\Psi}_\tau^{\text{P}} + {\boldsymbol{\Psi}}_{\tau}^{\text{S}} \commentTHg{=  {\boldsymbol{\Psi}}_{\tau}^{{\text{E}}_{\text{early}}} + \boldsymbol{\Psi}_\tau^{\text{I}}}
	\label{eq:psd_est_prior_error}
\end{equation}
with ${\boldsymbol{\Psi}}_{\tau}^{\text{P}} = {\boldsymbol{\Psi}}_{\tau}^{\text{D}_{\text{late}}} + {\boldsymbol{\Psi}}_{\tau}^{\text{N}}$ and \makebox{${\boldsymbol{\Psi}}_{\tau}^{{\text{E}}_{\text{early}}} = \mathbb{E}\left[ {\boldsymbol{e}}_{\text{early},\tau}^+ \left( {\boldsymbol{e}}_{\text{early},\tau}^+\right)^{\text{H}} \right]$}. We now analyze the dynamic behaviour of the different \acp{PSD}. The early echo error \ac{PSD} ${\boldsymbol{\Psi}}_{\tau}^{{\text{E}}_{\text{early}}}$ is assumed to be time-variant and its norm decreases during convergence of the \ac{AF} partitions $\hat{\boldsymbol{w}}_{b, \tau}$. In contrast, we can assume the \ac{PSD} matrix of the late echo and background noise ${\boldsymbol{\Psi}}_{\tau}^{\text{P}}={\boldsymbol{\Psi}}_\tau^{\text{D}_\text{late}} + {\boldsymbol{\Psi}}_\tau^{\text{N}}$ to be only slowly time-varying. This is motivated by the temporal smoothing effect resulting from the tails of \acp{RIR} \cite{kuttruff2016room} and the characteristics of many background noise signals, e.g., microphone noise or babble noise. On the other hand, the near-end speaker \ac{PSD} matrix ${\boldsymbol{\Psi}}_{\tau}^{\text{S}}$ is modelled to be potentially rapidly time-varying following the dynamics of speech signals. 
% On the other hand, the near-end speaker \ac{PSD} matrix ${\boldsymbol{\Psi}}_{\tau}^{\text{S}}$ is modelled to be potentially rapidly time-varying following the dynamics of speech signals.
%The goal of any noise \ac{PSD} estimator for a \ac{PBKF} is to robustly estimate ${\boldsymbol{\Psi}}_{\tau}^{I} = {\boldsymbol{\Psi}}_{\tau}^{P} + {\boldsymbol{\Psi}}_{\tau}^{S}$.
% TODO: not perfectly desribed
%\begin{equation}
%	{\boldsymbol{\Psi}}_{\tau}^{I} = {\boldsymbol{\Psi}}_{\tau}^{P} + {\boldsymbol{\Psi}}_{\tau}^{S}.
%	\label{eq:psd_est_tilde_n}
%\end{equation}
%

While all state-of-the-art approaches aim at a direct estimation of ${\boldsymbol{\Psi}}_{\tau}^{\text{I}}$ from the prior error signal $\boldsymbol{e}_\tau^+$, we propose the additive \commentTHb{observation} noise \ac{PSD} estimator
\begin{equation}
	\hat{\boldsymbol{\Psi}}_{\tau}^{\text{I}} = \hat{\boldsymbol{\Psi}}_{\tau}^{\text{P}} + \hat{\boldsymbol{\Psi}}_{\tau}^{\text{S}}
	\label{eq:psd_est_tilde_n}
\end{equation}
which allows to exploit the different time-variance of the statistics of $\boldsymbol{p}_\tau$ and $\boldsymbol{s}_\tau$.
%to exploit the varying stationarity properties of the different signal components
%
%to decompose the estimator into
%%
%\begin{align}
%% \left[{\boldsymbol{\Psi}}_{\tau}^{I}\right]_{mm} = \left[{\boldsymbol{\Psi}}_{\tau}^{P}\right]_{mm} + \left[{\boldsymbol{\Psi}}_{\tau}^{S}\right]_{mm},
%{\boldsymbol{\Psi}}_{\tau}^{I} = {\boldsymbol{\Psi}}_{\tau}^{P} + {\boldsymbol{\Psi}}_{\tau}^{S},
%\end{align}
%%
%with ${\boldsymbol{\Psi}}_{\tau}^{P}={\boldsymbol{\Psi}}_\tau^{d_\text{late}} + {\boldsymbol{\Psi}}_\tau^{N}$, to exploit the varying statistical signal properties of its underlying components, i.e., $\boldsymbol{d}_{\text{late}, \tau}$, $\boldsymbol{n}_{\tau}$ and $\boldsymbol{s}_{\tau}$. While we assume the late echo 
%\ac{PSD} ${\boldsymbol{\Psi}}_\tau^{d_\text{late}}$ and the background noise \ac{PSD} ${\boldsymbol{\Psi}}_\tau^{N}$ to be only slowly non-stationary, the near-end speaker \ac{PSD} ${\boldsymbol{\Psi}}_\tau^{S}$ is modelled as rapidly non-stationary process. This is motivated by the characteristic properties of \acp{RIR} tails, which exhibit ... \cite{abc}, and the non-stationary characteristic of speech signals. 
Considering \acp{PF} that are designed to extract the desired near-end speaker $\tilde{\boldsymbol{s}}_{\tau}$ with minimum distortion  from the prior error $\tilde{\boldsymbol{e}}_\tau^+$ (cf.~Eq.~\eqref{eq:cost_func}), a straightforward estimator for the near-end \ac{PSD} ${\boldsymbol{\Psi}}_{\tau}^{\text{S}}$ is given by the periodogram of the masked prior error
\begin{align}
	\left[\hat{\boldsymbol{\Psi}}_{\tau}^{\text{S}}\right]_{mm} =\lambda_{S} \left[\hat{\boldsymbol{\Psi}}_{\tau-1}^{\text{S}}\right]_{mm} + (1-\lambda_{S}) \left| \left[ \widehat{\boldsymbol{M}}_\tau \boldsymbol{e}_\tau^+  \right]_m \right|^2
	\label{eq:obs_noise_est_S}
\end{align}
with the recursive averaging factor $\lambda_{S}$. Note that due to the same \ac{DFT} length $M$ the estimated \ac{STFT} mask $\widehat{\boldsymbol{M}}_\tau$ can be similarly applied in the overlap-save domain. By subtracting the near-end \ac{PSD} matrix estimate $\hat{\boldsymbol{\Psi}}_{\tau}^{\text{S}}$ from Eq.~\eqref{eq:psd_est_prior_error} and assuming \makebox{${\boldsymbol{\Psi}}_{\tau}^{\text{S}} \approx \hat{\boldsymbol{\Psi}}_{\tau}^{\text{S}}$}, the prior error \ac{PSD} matrix is given by \makebox{${\boldsymbol{\Psi}}_{\tau}^{{\text{E}}} \approx {\boldsymbol{\Psi}}_{\tau}^{\text{E}_{\text{early}}} + \boldsymbol{\Psi}_\tau^{\text{P}}$}. As the late echo and background noise \ac{PSD} matrix ${\boldsymbol{\Psi}}_{\tau}^{\text{P}}$ varies only slowly compared to the early echo error \ac{PSD} matrix ${\boldsymbol{\Psi}}_{\tau}^{\text{E}_{\text{early}}}$, any stationary noise \ac{PSD} estimator can be used for its inference. In this paper we use the minimum statistics estimator \cite{minimum_statistics} which is given by the minimum of the latest $\kappa$ estimates
%
%early echo \ac{PSD} ${\boldsymbol{\Psi}}_{\tau}^{\tilde{E}}$ is vastly more non-sta
%
%In contrast to the near-end speaker signal, the late echo component ${\boldsymbol{\Psi}}_\tau^{d_\text{late}}$ and the background noise ${\boldsymbol{\Psi}}_\tau^{N}$ are assumed to be only slowly non-stationary. This can be justified by the late part of an \ac{RIR} being ...  \cite{abc}. Due to the similar stationarity characteristic of ${\boldsymbol{\Psi}}_\tau^{d_\text{late}}$ and ${\boldsymbol{\Psi}}_\tau^{N}$, we suggest to estimate directly the sum of the individual \acp{PSD}, i.e., ${\boldsymbol{\Psi}}_{\tau}^{P}={\boldsymbol{\Psi}}_\tau^{d_\text{late}} + {\boldsymbol{\Psi}}_\tau^{N}$. 
%
%\vspace*{1cm}
%A computationally efficient and robust estimation of non-statioanry
%
\begin{align}
	\left[{\hat{\boldsymbol{\Psi}}}_{\tau}^{\text{P}}\right]_{mm} = \min \begin{bmatrix}
		\left[{\boldsymbol{\Upsilon}}_{\tau-\kappa+1}^{\text{P}}\right]_{mm}, & \dots, & \left[{\boldsymbol{\Upsilon}}_{\tau}^{\text{P}}\right]_{mm} \\
	\end{bmatrix} 
	\label{eq:obs_noise_est_P}
\end{align}
of a smoothed periodogram 
\begin{equation}
	\left[{\boldsymbol{\Upsilon}}_{\tau}^{\text{P}}\right]_{mm} = \lambda_{P} \left[{\boldsymbol{\Upsilon}}_{\tau-1}^{\text{P}}\right]_{mm} + (1-\lambda_{P}) \left| \left[\hat{\boldsymbol{p}}_\tau \right]_m \right|^2
	\label{eq:obs_noise_est_P_tilde}
\end{equation}
with \commentTHf{the} late echo and background noise estimate \makebox{$\hat{\boldsymbol{p}}_\tau= \left(\boldsymbol{I}_M -\widehat{\boldsymbol{M}}_{\tau} \right) \boldsymbol{e}_\tau^+$} and the recursive averaging factor $\lambda_P$. Note that ${\hat{\boldsymbol{\Psi}}}_{\tau}^{\text{P}}$ can be interpreted as a temporally smoothly changing minimum regularization in the \ac{KF} step size \eqref{eq:eStep}.
%
%
%$\boldsymbol{p}_\tau=\boldsymbol{d}_{\text{late},\tau} + \boldsymbol{n}_\tau$
%
%
%Discuss that due to non-statioarny speaker is often zero and this would results in badly conditioned step sizes
%
%thus minimum step size 
%
%can be interpreted as minimum regularitzation
%
%\begin{align}
%\left[{\boldsymbol{\Psi}}_{\tau}^{I}\right]_{mm} = \left[{\boldsymbol{\Psi}}_{\tau}^{S}\right]_{mm} + \left[{\boldsymbol{\Psi}}_{\tau}^{D_{\text{late},N}}\right]_{mm}
%\end{align}
%
%
Finally, \commentTHg{the process noise \commentTHb{\ac{PSD} matrices are estimated} by \cite{kuech_state-space_2014}}
% Finally, as proposed in \cite{kuech_state-space_2014} the process noise \commentTHb{\ac{PSD} matrices are estimated} by
%
%\begin{align}
%%\boldsymbol{\Psi}_{t,l}^S &= {\e}^+_\tau {\e}^{+ \herm}_\tau + {\C}_\tau {\bP}_{\tau} {\C}_\tau \\
%\left[ \hat{\boldsymbol{\Psi}}_{b,\tau}^{\text{W}} \right]_{mm} = (1-A^2)~   \mathbb{E}\left[ \left[ \hat{\w}_{b,\tau} {\hat{\w}_{b,\tau}}^{\text{H}} \right]_{mm} \right]	\label{eq:mStepStateNoise}
%\end{align}
\begin{align}
	% \left[ \hat{\boldsymbol{\Psi}}_{b,\tau}^{\Delta\text{W}} \right]_{mm} = (1-A^2)~ \left[ \lambda_W \commentTHa{\hat{\boldsymbol{\Psi}}_{b,\tau-1}^{\Delta\text{W}}}  + (1-\lambda_W) \hat{\w}_{b,\tau} {\hat{\w}_{b,\tau}}^{\text{H}}  \right]_{mm}. \label{eq:mStepStateNoise}
	\left[ \hat{\boldsymbol{\Psi}}_{b,\tau}^{\Delta\text{W}} \right]_{mm} = (1-A^2)~\left[ \hat{\boldsymbol{\Psi}}_{b,\tau}^{\text{W}}  \right]_{mm} \label{eq:mStepStateNoise}
\end{align}
with \commentTHf{$\hat{\boldsymbol{\Psi}}_{b,\tau}^{\text{W}} = \lambda_W {\hat{\boldsymbol{\Psi}}_{b,\tau-1}^{\text{W}}}  + (1-\lambda_W) \hat{\w}_{b,\tau-1} {\hat{\w}_{b,\tau-1}}^{\text{H}}$}.

\section{Algorithmic Description}
\label{sec:alg_descr}
The proposed echo cancellation scheme for one block of \commentTHb{microphone samples} $\boldsymbol{y}_{\tau}$ is illustrated in Fig.~\ref{fig:alg_overview} and summarized in Alg.~\ref{alg:prop_alg_descr}. After computing the prior error $\boldsymbol{e}_\tau^+$, using the previous \ac{AF} estimate (cf.~Eq.~\eqref{eq:eStep}), the mask $\widehat{\boldsymbol{M}}_\tau$ is inferred by the \ac{DPF} (cf.~Sec.~\ref{sec:deep_pf}).
Subsequently, \commentTHg{the \ac{PSD} {matrices} $\boldsymbol{\Psi}_{b,\tau}^{\Delta\text{W}}$ and $\boldsymbol{\Psi}_\tau^{\text{I}}$} are estimated by Eqs.~\eqref{eq:mStepStateNoise} and \eqref{eq:psd_est_tilde_n}, respectively. 
%Subsequently, {the process noise \ac{PSD} \commentTHb{matrices} $\boldsymbol{\Psi}_{b,\tau}^{\Delta\text{W}}$ and \commentTHb{the} observation noise \ac{PSD} matrix $\boldsymbol{\Psi}_\tau^{\text{I}}$} are estimated by Eqs.~\eqref{eq:mStepStateNoise} and \eqref{eq:psd_est_tilde_n}, respectively. 
Note that if the initial matrices are chosen to be diagonal, the estimators ensure all subsequent estimates to be diagonal as well. Afterwards, the \commentTHb{means $\hat{\boldsymbol{w}}_{b,\tau}$ and state uncertainty matrices $\boldsymbol{P}_{b,\tau}$} of the \ac{AF} partitions $\boldsymbol{w}_{b,\tau}$ are updated by the \ac{KF} (cf.~Eq.~\eqref{eq:eStep}) using the estimated \ac{PSD} matrices $\hat{\boldsymbol{\Psi}}_{\tau}^{\text{I}}$ and $\hat{\boldsymbol{\Psi}}_{b,\tau}^{\Delta\text{W}}$. Finally, the time-domain near-end signal $\hat{\underline{\boldsymbol{s}}}_\tau$ is computed by applying the inverse \ac{STFT} to the \ac{DPF} estimate $\hat{\tilde{\boldsymbol{s}}}_\tau$.

\begin{algorithm}[tb] %[tb]
	%\caption{Supervised Adaptive Filtering by Local Projection-Based Gradient Denoising} % (\textcolor{red}{Check that this is exactly what is implemented in algorithm})
	\caption{Proposed \ac{KF}+\ac{DPF} algorithm for one block of microphone samples.} % \ac{OSASI} by {SSFDAF-NMF-EM-$L$}
	\label{alg:prop_alg_descr}
	\begin{algorithmic}
		%\For{$\tau=1,\dots,T$}
		\State Compute prior error $\boldsymbol{e}_\tau^+$ (cf.~Eq.~\eqref{eq:eStep})
		\State Infer \ac{DPF} mask $\widehat{\boldsymbol{M}}_\tau$ (cf.~Sec.~\ref{sec:deep_pf})
		\State \commentTHb{Update} \ac{PSD} \commentTHb{estimates} $\hat{\boldsymbol{\Psi}}^{\text{I}}_{\tau}$, $\hat{\boldsymbol{\Psi}}^{\Delta\text{W}}_{b,\tau}$ (cf.~Eqs.~\eqref{eq:psd_est_tilde_n}~and~\eqref{eq:mStepStateNoise})
		\State Update \ac{AF} estimates $\hat{\boldsymbol{w}}_{b,\tau }, {\bP}_{b,\tau} $ (cf.~Eq.~\eqref{eq:eStep})
		\State Compute time-domain near-end signal $\underline{\hat{\boldsymbol{s}}}_\tau$ by inverse \ac{STFT}
		%\EndFor
	\end{algorithmic}
\end{algorithm}

%\clearpage
%\newpage

\section{Experiments}
\label{sec:experiments}
In this section, the proposed algorithm is evaluated for a large variety of \ac{AEC} scenarios. Each scenario is created by randomly drawing a far-end and a near-end speech signal from the \textit{LibriSpeech} database \cite{7178964} comprising $283$ different speakers. Subsequently, the clean echo signal \commentTHb{$\underline{\boldsymbol{d}}_\tau$} is simulated by convolving the far-end signal with a randomly drawn \ac{RIR} from the databases \cite{jeub_binaural_2009,Wen06evaluationof, mird}, comprising $201$ different \acp{RIR} \commentTHg{with reverberation times \makebox{$T_{60}$} ranging from \makebox{$120$ ms} to \makebox{$780$ ms}}. Finally, the near-end speaker signal and white Gaussian sensor noise are added. %
\commentTHg{Both signals are scaled according to a random near-end-to-echo and {\acl{ENR}} in the ranges \makebox{$\left[-10, ~10\right]$ dB} and \makebox{$\left[30,~ 35\right]$ dB, respectively.}}
%Both signals are scaled according to a random \commentTHb{\acl{NER}} in between  \makebox{$-10$ dB} and \makebox{$10$ dB} and a random \commentTHb{\acl{ENR}} in between  \makebox{$30$ dB} and \makebox{$35$ dB}. 
%\commentTHg{An \ac{EPC} is simulated by randomly drawing \acp{RIR} and signals for simulating the observations before and after a random switching time in between \makebox{$7.2$ s} and \makebox{$8.8$ s}. }
\commentTHj{An \ac{EPC} is simulated by randomly drawing \acp{RIR} and signals for simulating the observations before and after a specific switching time. The switching time was chosen randomly in between \makebox{$7.2$ s} and \makebox{$8.8$ s} to avoid overfitting of the \ac{DNN} to a fixed point in time.}
% \textcolor{red}{An \ac{EPC} is simulated by randomly drawing \acp{RIR} and signals for simulating the observations before and after a specific switching time. The switching time was chosen randomly in between \makebox{$7.2$ s} and \makebox{$8.8$ s} to avoid overfitting of the \ac{DNN}.}
% \commentTHj{An \ac{EPC} is simulated by randomly drawing \acp{RIR} and signals for simulating the observations before and after a random switching time. The switching time was chosen randomly in between \makebox{$7.2$ s} and \makebox{$8.8$ s} to avoid overfitting of the \ac{DNN}.}
% An \ac{EPC} is simulated by randomly drawing a signal duration in between \makebox{$7.2$ s} and \makebox{$8.8$} s and afterwards appending a new scenario. 
%In total approximately $5$ hours of data was generated which was split into \makebox{$4.5$} hours of training data and $27$ minutes of disjoint, i.e, different speakers and \acp{RIR}, testing data.
% The block shift was set to $R=256$ samples with a sampling frequency of \makebox{$f_s=16$ kHz}. The \ac{PBKF} modelled \commentTHa{$B=8$ partitions which corresponds to an overall filter length of $L=2048$ samples, i.e., \makebox{$62.5$ ms}.} 
%
The block shift was set to $R=256$ samples with a sampling frequency of \makebox{$f_s=16$ kHz}. The \ac{PBKF} modelled \commentTHa{$B=8$ partitions which corresponds \commentTHg{to a filter length} of $L=2048$ samples.} The noise \ac{PSD} estimators used the \commentTHb{parameters} $\commentTHa{\lambda_{P}=\lambda_W=0.9}$, $\lambda_{S}=0$ \commentTHg{and $\kappa =90$}. \commentTHa{The input features to the \ac{DPF} were computed by using a Hamming window and \makebox{$\epsilon_1 = 10^{-12}$}.} The \ac{DPF} \commentTHb{used} approximately $3.4$ million parameters with the input dimension of the stacked \commentTHg{\ac{GRU}} layers being $P=512$. \commentTHa{It} was trained using the ADAM optimizer \commentTHb{\cite{kingma2014adam}} with a step size of \commentTHb{$10^{-3}$}\commentTHa{, a regularization factor of \makebox{$\epsilon_2=10^{-12}$}} and $4.4$ hours of training data. The training data was preprocessed by a {\ac{PBKF}} for which the required \ac{PSD} matrices ${\boldsymbol{\Psi}}_{\tau}^{\text{S}}$ and $ {\boldsymbol{\Psi}}_{b,\tau}^{\Delta\text{W}}$ were estimated by Eqs.~\eqref{eq:psd_est_tilde_n} - \eqref{eq:mStepStateNoise}. Here, the estimated mask $\widehat{\boldsymbol{M}}_\tau$ was replaced by an oracle mask. The testing data (27 mins) was disjoint from the training data, i.e., different speakers and \acp{RIR}.
%
%
% and \makebox{$4.4$} hours of training data. The testing data consisted of $27$ minutes and was disjoint from the training data, i.e., different speakers and \acp{RIR}.
%
%The speakers and \acp{RIR} for creating the training data were disjoint from the evaluation data and split in a ratio of $9/1$.
%
\begin{figure}[t!]
	%\vspace*{-1cm}
	\centering
	\newlength\fwidth
	\setlength\fwidth{.98\columnwidth}
	\hspace*{-.15cm}\input{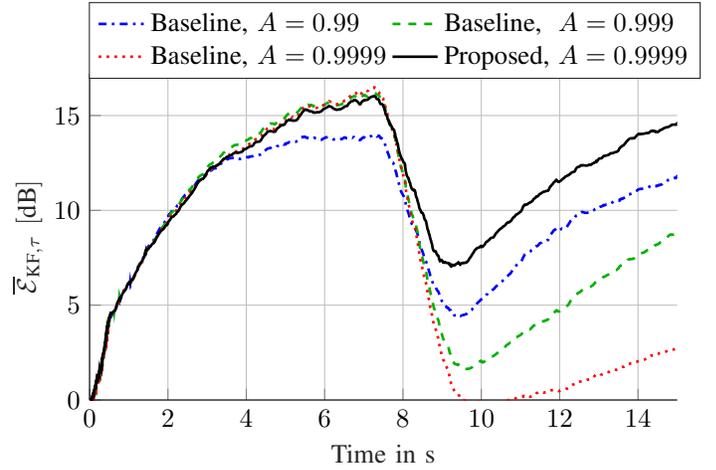}
	% \input{images/temp_erle_more_A.tikz}
	%\vspace*{-.6cm}
	\caption{Time-dependent ERLE of the {\ac{PBKF}} for various parametrizations of the baseline noise \ac{PSD} estimators and the proposed estimator.}
	\label{fig:temp_erle}
\end{figure}

\commentTHb{In the first experiment} the reconvergence behaviour of the {\ac{PBKF}} after abrupt \acp{EPC} \commentTHb{is} compared for the proposed \ac{PSD} estimator (cf.~Sec.~\ref{sec:psd_est}) relative to the state-of-the-art approach \cite{franzen_improved_2019}, i.e., recursively averaging the prior error \commentTHb{power $|\left[\boldsymbol{e}_{\tau}^+\right]_m |^2$} with an averaging factor of $0.5$. We compared several choices for the state transition parameter $A$ because \cite{yang_frequency-domain_2017} reports a trade-off between steady-state performance and reconvergence behaviour. \commentTHa{As performance measure} we used the time-dependent logarithmic \ac{ERLE}
% As objective performance measure we used the time-dependent logarithmic \ac{ERLE} of the prior error without the background noise and near-end speaker interference
%
% after the \ac{PBKF}
%
\begin{align}
	%{\mathcal{E}}_{\text{KF},\tau}  = g(\mathbb{E}\left[||\underline{\boldsymbol{d}}_{\tau}||_2^2\right], \mathbb{E}\left[||\underline{\boldsymbol{d}}_{\tau}-\hat{\boldsymbol{\underline{d}}}_{\tau}||_2^2\right]),
	{\mathcal{E}}_{\text{KF},\tau} = 10 \log_{10} \frac{\mathbb{E}\left[||\underline{\boldsymbol{d}}_{\tau}||^2\right] }{\mathbb{E}\left[||\underline{\boldsymbol{d}}_{\tau}-\hat{\boldsymbol{\underline{d}}}_{\tau}||^2\right]},
\end{align}
%
%with $g(\boldsymbol{{a}}, \boldsymbol{{b}}) =  10 \log_{10} \left(\frac{||\boldsymbol{{a}}||^2}{||\boldsymbol{{b}}||^2} \right)$ and 
with $||\cdot||^2$ denoting the squared Euclidean norm and the expectation being approximated by temporal recursive averaging. To allow for more general conclusions, the time-dependent logarithmic \ac{ERLE} ${\mathcal{E}}_{\text{KF},\tau}$ has been averaged over $100$ different scenarios. The resulting average time-dependent \ac{ERLE} $\overline{{\mathcal{E}}}_{\text{KF},\tau}$ for various choices of the state transition parameter $A$ for the baseline and the proposed noise \ac{PSD} estimator is shown in Fig.~\ref{fig:temp_erle}. As can be concluded from Fig.~\ref{fig:temp_erle} for the baseline, larger state transition parameters $A$ result in better steady-state performance at the cost of slower reconvergence after \acp{EPC} due to an overestimation of the noise \ac{PSD}. The proposed noise \ac{PSD} estimator, however, avoids this trade-off and allows for both high steady-state performance \commentTHa{and rapid reconvergence.}
% The proposed noise \ac{PSD} estimator, however, avoids this trade-off and allows for both high steady-state performance and a rapid reconvergence after the \ac{EPC}.
%

Finally, we evaluate the echo suppression and near-end distortion performance of the individual algorithmic components, i.e., {\ac{PBKF}}-only and \ac{DPF}-only, and their synergistic combination \ac{PBKF}+\ac{DPF}. As performance measures we use:
%
%\begin{align}
%\label{eq:sigPowRatio}
%\begin{aligned}
%{\mathcal{E}}_{\text{KF}} &= g(\underline{\boldsymbol{d}} ,\underline{\boldsymbol{d}} - \hat{\underline{\boldsymbol{d}}}), \\   {\mathcal{E}}_{\text{PF}} &= g(\underline{\boldsymbol{d}}, \text{pf}(\underline{\boldsymbol{d}} - \hat{\underline{\boldsymbol{d}}}))    ,  
%\end{aligned}
%&&
%\begin{aligned}
% {\mathcal{S}}_{\text{pf}} &= g(\beta \underline{\boldsymbol{s}},\beta \underline{\boldsymbol{s}} - \text{pf}(\underline{\boldsymbol{s}})) \\ {\Delta \Upsilon} &= \text{PESQ}(\underline{\boldsymbol{s}}, \hat{\underline{\boldsymbol{s}}}) - \text{PESQ}({\underline{\boldsymbol{s}}}, \underline{\boldsymbol{y}}) ,
%\end{aligned}
%\end{align}
%
%\begin{align}
%\label{eq:sigPowRatio}
%\begin{aligned}
%&{\mathcal{E}}_{\text{KF}} = 10 \log_{10} \frac{||\underline{\boldsymbol{d}}||^2}{||\underline{\boldsymbol{d}} - \hat{\underline{\boldsymbol{d}}}||^2}, \\   &{\mathcal{E}}_{\text{PF}} = 10 \log_{10} \frac{||\underline{\boldsymbol{d}}||^2}{|| \text{pf}(\underline{\boldsymbol{d}} - \hat{\underline{\boldsymbol{d}}})||^2}   ,  
%\end{aligned}
%&&
%\begin{aligned}
%&{\mathcal{S}}_{\text{PF}} = 10 \log_{10} \frac{||\beta \underline{\boldsymbol{s}}||^2}{||\beta \underline{\boldsymbol{s}} - \text{pf}(\underline{\boldsymbol{s}})||^2} \\ 
%&{ \Delta\text{PESQ}} = \text{pq}(\underline{\boldsymbol{s}}, \underline{\hat{\boldsymbol{s}}}) - \text{pq}({\underline{\boldsymbol{s}}}, \underline{\boldsymbol{y}}) \vphantom{\frac{||}{||}}.
%\end{aligned}
%\end{align}
%
\begin{align}
	%\label{eq:sigPowRatio}
	& {\mathcal{E}}_{\text{PF}} = 10 \log_{10} \frac{||\underline{\boldsymbol{d}}||^2}{|| \text{pf}(\underline{\boldsymbol{d}} - \widehat{\underline{\boldsymbol{d}}})||^2}   , ~
	{\mathcal{S}}_{\text{PF}} = 10 \log_{10} \frac{||\beta \underline{\boldsymbol{s}}||^2}{||\beta \underline{\boldsymbol{s}} - \text{pf}(\underline{\boldsymbol{s}})||^2}, \notag \\
	&{\mathcal{E}}_{\text{KF}} = 10 \log_{10} \frac{||\underline{\boldsymbol{d}}||^2}{||\underline{\boldsymbol{d}} - \widehat{\underline{\boldsymbol{d}}}||^2}, ~~
	{ \Delta\text{PESQ}} = \text{pq}(\underline{\boldsymbol{s}}, \underline{\hat{\boldsymbol{s}}}) - \text{pq}({\underline{\boldsymbol{s}}}, \underline{\boldsymbol{y}}) \vphantom{\frac{||}{||}}. \notag
\end{align}
%
%\begin{align}
%\label{eq:sigPowRatio}
%&{\mathcal{S}}_{\text{PF}} = 10 \log_{10} \frac{||\beta \underline{\boldsymbol{s}}||^2}{||\beta \underline{\boldsymbol{s}} - \text{pf}(\underline{\boldsymbol{s}})||^2}, 
%&{\mathcal{E}}_{\text{PF}} = 10 \log_{10} \frac{||\underline{\boldsymbol{d}}||^2}{|| \text{pf}(\underline{\boldsymbol{d}} - \hat{\underline{\boldsymbol{d}}})||^2}   , \\  
%&{\mathcal{E}}_{\text{KF}} = 10 \log_{10} \frac{||\underline{\boldsymbol{d}}||^2}{||\underline{\boldsymbol{d}} - \hat{\underline{\boldsymbol{d}}}||^2},  
%&{ \Delta\text{PESQ}} = \text{pq}(\underline{\boldsymbol{s}}, \underline{\hat{\boldsymbol{s}}}) - \text{pq}({\underline{\boldsymbol{s}}}, \underline{\boldsymbol{y}}) \vphantom{\frac{||}{||}}.
%\end{align}
%
\begin{table}[t]
	%\vspace*{-.7cm}
	\caption{Mean and standard deviation (in parentheses) of the various components of the proposed \ac{AEC} algorithm. For the best performance values bold font is used.}\vspace{-0mm}
	% Performance evaluation of the various algorithmic parts for the proposed algorithm. 
	\vspace*{-.05cm}
	\setlength{\tabcolsep}{10.0pt}
	\begin{center}
		\begin{tabular}{c c c c}
			\toprule
			& {PBKF-only} &\commentTHg{DPF}-only& \commentTHg{PBKF+DPF}  \\ \midrule
			$t_\text{pr}$[ms] / RTF & $\phantom{0}\textbf{0.4}/\textbf{0.03}$ &  $\phantom{0}1.0/0.06$ & $\phantom{0}1.4/0.09$ \\ 
			$\overline{\mathcal{E}}_{(\cdot)}$ & $10.5~(2.7)$ &   $13.2~(2.8)$ & $\textbf{17.0}~(3.6)$  \\ 
			$\overline{\mathcal{S}}_{\text{PF}}$ & $\phantom{0}\infty$ & $14.6~(3.2)$ & $\textbf{26.4}~(5.7)$  \\ 
			$ \overline{\Delta\text{PESQ}}$ & $\phantom{0}0.55~(0.4)$& $\phantom{0}0.67~(0.4)$ & $\textbf{1.12}~(0.5)$ \\
			\bottomrule 
		\end{tabular} 
		%\vspace{-.4cm}
	\end{center}
	\label{tab:tabPerfMeas}
	\vspace{-.3cm}
\end{table}
%\vspace*{-.2cm}
%
Here, \commentTHb{the \ac{ERLE} averaged over the entire signal duration}, denoted by omitting the time index $\tau$, after the {\ac{PBKF}} and the \ac{DPF} is represented by $\mathcal{E}_{\text{KF}}$ and $\mathcal{E}_{\text{PF}}$, respectively. The near-end distortion is measured by $\mathcal{S}_{\text{PF}}$, with the scaling factor \makebox{$\beta = \frac{\boldsymbol{{\boldsymbol{s}}}^{\text{T}} \text{pf}(\boldsymbol{s})}{||\boldsymbol{{s}}||^2}$} \cite{roux_sdr_2019}, and the PESQ (Perceptual Evaluation of Speech Quality \cite{pesq}) improvement $\Delta \text{PESQ}$. Note that the processing of a signal by the \ac{DPF} is described by $\text{pf}(\cdot)$ while $\text{pq}(\cdot, \cdot)$ denotes the computation of the {PESQ} \cite{pesq}. Tab.~\ref{tab:tabPerfMeas} shows the arithmetic averages of $100$ random experiments of \commentTHa{the performance measures} ${\mathcal{E}}_{(\cdot)}$, ${\mathcal{S}}_{\text{PF}}$ and ${\Delta\text{PESQ}}$, denoted by an overbar. Furthermore, the runtime  $t_{\text{pr}}$ to process one signal block on an \textit{\commentTHa{Intel Xeon CPU E3-1275 v6 @ 3.80GHz}} and the corresponding \ac{RTF} are given. Note that only the proposed noise \ac{PSD} estimator (cf.~Sec.~\ref{sec:psd_est}) is evaluated due to the limited reconvergence capabilities of the baseline estimator. 
% Furthermore, to show the effect \commentTHg{of} the \ac{PBKF}, we evaluated a \ac{DPF}-only algorithm which was trained with the microphone signal \commentTHg{$\underline{\boldsymbol{y}}_{\tau}$} instead of the error signal \commentTHg{$\underline{\boldsymbol{e}}_{\tau}^+$} in the feature computation in Eq.~\eqref{eq:feature_e_x}. 
Furthermore, to show the effect {of} the \ac{PBKF}, we evaluated a \ac{DPF}-only algorithm which was trained with the microphone signal {$\underline{\boldsymbol{y}}_{\tau}$} instead of the error signal {$\underline{\boldsymbol{e}}_{\tau}^+$} in the \commentTHj{feature computation \eqref{eq:feature_e_x}.} We conclude from Tab.~\ref{tab:tabPerfMeas} that while using only a \ac{PBKF} results in limited echo cancellation, the \ac{DPF}-only approach introduced significant distortions. In contrast the combination allows for high echo attenuation while introducing only little distortions. 
%Furthermore, Tab.~\ref{tab:tabPerfMeas} shows the runtime $t_{\text{pr}}$ of the various algorithmic parts to process one signal block on an \textit{Intel(R) Xeon(R) CPU E3-1275 v6 @ 3.80GHz} and the corresponding real-time factor $\text{RTF}$. 
\commentTHa{Finally, we see from Tab.~\ref{tab:tabPerfMeas} that due to the modest computational requirements, the proposed method is well suited for real-time applications.}

\section{Conclusion}
\label{sec:summaryOutlook}
% In this paper, we proposed a novel synergistic \ac{KF}+\ac{DPF} algorithm which outperforms known state-of-the-art \ac{AEC} algorithms for time-varying acoustic scenarios in terms of reconvergence speed without compromising performance for static scenarios. This is achieved by efficiently exploiting the \ac{DPF} near-end estimate and the \ac{KF} estimation error. Without any auxiliary mechanisms, we thereby overcome the limitations of \ac{KF}-based step-size adaptation algorithms.
In this paper, we proposed a novel synergistic \ac{KF}+\ac{DPF} algorithm which \commentTHa{improves state-of-the-art} \ac{AEC} algorithms for time-varying acoustic scenarios in terms of reconvergence speed without compromising performance for static scenarios. This is achieved by efficiently exploiting the \ac{DPF} near-end estimate and the \ac{KF} estimation error \commentTHa{for \commentTHb{inferring} an \ac{AF} step-size.} Without any auxiliary mechanisms, we thereby overcome the limitations of \ac{KF}-based step-size adaptation algorithms.

\bibliographystyle{IEEEbib}
{\small\bibliography{refs}}
% {\small	\bibliography{refs}}

%\newpage
%\clearpage
%
%\section{\textcolor{red}{TODO}}
%
%\begin{itemize}
%	\item cite your own ICASSP NMF publication \cite{kfNMF}
%\end{itemize}

\end{document}